# CMS Sematrix: A Tool to Aid the Development of Clinical Quality Measures (CQMs)

Michael A. Schwemmer[1], Po-Hsu Chen[1], Mithun Balakrishna[2], Amy Leibrand[3], Aaron Leonard[2], Nancy J. McMillan[1], and Jeffrey J. Geppert[3]

[1] Advanced Analytics, Battelle Memorial Institute, Columbus, OH, USA

[2] Lymba Corporation, Richardson, TX USA

[3] Health Research, Battelle Memorial Institute, Columbus, OH, USA

## Summary

**Background:** As part of the effort to improve quality and to reduce national healthcare costs, the Centers for Medicare & Medicaid Services (CMS) are responsible for creating and maintaining an array of clinical quality measures (CQMs) for assessing healthcare structure, process, outcome, and patient experience across various conditions, clinical specialties, and settings. The development and maintenance of CQMs involves substantial and ongoing evaluation of the evidence on the measure's properties—importance, reliability, validity, feasibility, and usability. As such, CMS conducts monthly environmental scans of the published clinical and health services literature. Conducting time consuming, exhaustive evaluations of the ever-changing healthcare literature presents one of the largest challenges to an evidence-based approach to healthcare quality improvement. Thus, it is imperative to leverage automated techniques to aid CMS in the identification of clinical and health services literature relevant to CQMs. Additionally, the estimated labor hours and related cost savings of using CMS Sematrix compared to a traditional literature review are roughly 818 hours and $122,000 for a single monthly environmental scan [1].

**Objective:** Designing CMS Sematrix, an automated knowledge extraction framework that scans published clinical and health services literature, identifies relevant articles for a given CQM, and stores evidence presented by the articles in a form capable of analysis and synthesis.

**Methods:** CMS Sematrix contains three major components: (1) a quality measure ontology to describe high-level knowledge constructs contained in CQM; (2) a natural language process (NLP) system to extract concepts and relations that correspond to the ontology from text; and (3) a graphical database to store the concepts and relations extracted from text as Resource Description Framework (RDF) triples. To build the framework, a set of 65 CQMs covering a variety of healthcare domains and 98 biomedical articles (PubMed Abstracts and PubMed Central Full Articles) were manually annotated with CQM ontology specific concepts and relations. In addition, the 65 CQMs were manually reviewed by subject matter experts in order to extract the high-level quality constructs. Lastly, to validate that the documents returned by CMS Sematrix contain information relevant to the given quality measure, we developed an automated procedure for identifying relevant documents. The results of this automated procedure were then compared to a manual document review for set of 9 randomly selected measures from the set of 65 CQMs.

**Results:** The NLP component of CMS Sematrix was able to correctly identify CQM concepts with an average recall score of 87% for measure descriptions and 86% for articles. In addition, CMS Sematrix achieved overall precision and recall scores of 84% and 62% when extracting concept relations. We then conducted an environmental scan of the PubMed and PubMed Central abstracts and articles using the set of 65 CQMs. For the 9 measures selected for manual review, our automated procedure for determining relevant documents obtained average precision and recall scores of 84% and 88%. Running this procedure on the full set of 65 CQMs, we found that on average roughly 72% of the articles returned by CMS Sematrix for a given measure contain information relevant to the measure description using our June 2018 environmental scan data.

**Conclusions:** CMS Sematrix is able to identify articles published in the clinical and health services literature that contain information relevant to a given CQM. In practice, CMS Sematrix can reduce the time-consuming burden of the CMS monthly environmental scans and allow measure developers to quickly and accurately design CQMs to track outcomes in order to improve the national healthcare system.

# Introduction

The IMPACT Act [2], MACRA[3], and the 21st Century Cures Act [4] are three of the more recent legislative manifestations of the acknowledged importance of reducing the cost of healthcare while improving quality and enabling innovation. Recent estimates project that the cost of healthcare will reach nearly 20% of the Gross Domestic Product by 2026 [5], and those are dollars that might otherwise be spent on complementary societal needs like infrastructure, education, housing, and many others, especially at the state and local level. As the nation's largest payer for healthcare, the Centers for Medicare & Medicaid Services (CMS) is a central component to the success of this effort.

The transformation of the healthcare system from volume to value is the preferred mechanism to achieve cost reduction, quality improvement, and innovation. This transformation requires an array of clinical quality measures (CQMs) for assessing healthcare structure, process, outcome, and patient experience across various conditions, clinical specialties, and settings.  To achieve its quality and transformation priorities, CMS maintains an inventory of over 2,000 CQMs for use in quality improvement, comparative reporting, value-based purchasing, and alternative payment models (http://cmit.cms.gov).  The development and maintenance of CQMs involves substantial and ongoing evaluation of the evidence on the measure's properties—importance, reliability, validity, feasibility, and usability. The use of measures with poor reliability and validity wastes time and resources and may result in unintended system harms. Measures that are not feasible impose significant burden on consumers and clinicians. Measures must have information value to be usable for selecting clinicians or health plans (for consumers), allocating resources to quality improvement (for clinicians), or prioritizing clinical and health services research (for government).

To ensure this evidence is timely and complete, CMS conducts a monthly environmental scan of the published clinical and health services literature for all 2,000 CQMs. Conducting a scan for such a high volume of measures would be challenging enough; however, the challenge is further exacerbated with the rapid increase in the number of research publications. In 2017 alone, MEDLINE, the U.S. National Library of Medicine's database of journal articles on biomedicine, added more than 813,500 new citations [6]. Human review of the results of this scan would be cost prohibitive and would not keep pace with the increase in the number of publications. A human reviewer, no matter how proficient, must select relevant keywords to perform the search, read each returned abstract to establish relevance (or not) with the measure under consideration, rank the relevant abstracts to identify the subset of full-text articles to review, read the identified full-text articles, extract the knowledge contained in the full-text articles that provides evidence on the measure properties, and store that evidence in some form capable of analysis and synthesis. For a small set of measures in a common domain, a human review may take a 1,000 hours and cost hundreds of thousands of dollars.

To facilitate the monthly environmental scan for every measure in the CMS Measure Inventory, we have collaborated with the CMS Measures Management System (MMS) to develop a system called CMS Sematrix that automates the identification of clinical and health services literature relevant to CQMs, the extraction of knowledge contained in the relevant abstracts and full-text articles that provides evidence on the measure properties, and the store of that evidence in a form capable of analysis and synthesis. CMS Sematrix contains three major components: (1) A quality measure ontology to describe high-level knowledge constructs contained in CQM; (2) a natural language process (NLP) system to extract concepts and relations that correspond to the ontology from text; and (3) a graphical database to store the concepts and relations extracted from text as Resource Description Framework (RDF) [7] triples that can be queried to deduce measure components within documents. To our knowledge, there is no currently available off-the-shelf computational cognitive service that provides a competitive option to

CMS Sematrix due to its utilization of a highly specific clinical quality measure ontology created explicitly for use in our system.

## Objectives

The overall objective of this study is to design an automated knowledge extraction framework we call CMS Sematrix that scans the published clinical and health services literature, identifies relevant articles for a given CQM, and stores the evidence contained within the articles in a form capable of analysis and synthesis. To achieve this objective, we detail the steps required to build the individual components that make up the CMS Sematrix system. Namely, the definitions of the CQM ontology, the structure and training methodology of the NLP engine, and the resulting knowledge database. Lastly, we aim to show that CMS Sematrix dramatically reduces the labor hours and related cost compared to a traditional literature review without losing much accuracy for developing and maintaining CQMs, and the results returned by the system are relevant to CQM developers.

## Methods

### Clinical Quality Measure (CQM) Ontology

The goal of the CQM ontology is to standardize the essential features of a CQM into a set of abstract concepts with defined relationships between them. The components of the measure, such as the measure focus, target population, quality construct, and quality priority, can then be systematically represented as combinations of these concepts. This allows NLP tools to identify and extract these concepts and relations and place them in a structured format that can be used for semantic reasoning and analysis. The specific application for the CMS was to extract these concepts and relations from both clinical and health services research articles and the measure description text to identify articles that contain information relevant to a specific measure.

The abstract concepts in the ontology are displayed in Table 1. It is important to note that the Population concept can also have the attributes "Age Group", "Gender", or social determinants of health which can be used for further refinement. Similarly, the health status concept has attributes "severity" and "time". In addition to concepts, we have defined the ways in which the concepts can relate to each other (see Table 2). Each relation has a specified domain and range among the concepts, as denoted in the table.

**Table 1. Abstract CQM Concepts.** The five high-level measure concepts captured by the CQM ontology, along with their definitions and examples.

| Concept | Definition | Examples |
| --- | --- | --- |
| Change Concept | Healthcare activities that increase the likelihood of desired health outcomes | Medication, Screening, Surgery |
| Health Status | Signs or Symptoms, disorder, disease, complication, functional status, advanced illness | Acute myocardial infarction, Diabetes, Glaucoma |
| Population | Population and related concepts | Patients, Cohort |
| Utilization | Use of health care services | Hospital, Outpatient, Intensive care unit |
| Output | Outcome of interest | Reduce, Decrease, Improve |

**Table 2. CQM Relations with Domain and Range.** The five base semantic relations in the CQM ontology along with their definitions and the concepts they relate.

| Relation | Definition | (Domain, Range) |
| --- | --- | --- |
| Experiences | A particular instance of personally encountering or undergoing something | (Population, Change Concept) |
| HasFocus | To direct one's attention or efforts | (Change Concept, Health Status) |
| IsAPartOf | Represents how objects combine to form composite objects | (Health Status, Output) (Utilization, Output) (Utilization, Change Concept) |
| IsMadeUpOf | Represents how objects combine to form composite objects | (Population, Health Status) (Population, Utilization) |
| ResultsIn | To spring, arise, or proceed because of actions, circumstances, premises, etc.; be the outcome | (Change Concept, Output) |

## Natural Language Processing (NLP)

The recent emergence of Big Data- RDF triple stores makes it possible to merge massive amounts of structured and unstructured data by defining a common ontology model for representing the domain knowledge and storing all the domain assertions as semantic triples. However, technology gaps exist. More specifically, there is a lack of: (1) efficient and accurate algorithms or tools to automatically transform unstructured document content into knowledge graphs; (2) rich and complete semantic representation to store and query actionable domain knowledge that is compatible with the RDF standard; and (3) methods for accessing information to enable intelligent search applications while hiding the underlying complexity of the voluminous semantic data being searched.

CMS Sematrix uses the K-Extractor [8] NLP technology for the extraction of detailed semantic statements from unstructured text. The driver of the K-Extractor is the deep NLP Pipeline (Figure 1), which spans the lexical, syntactic, and semantic layers of knowledge extraction from text. It acts as a pipeline for filtering, data reduction, and value-added (semantics) functions, and discovers concepts and relations relevant to the ontology in the form of entities and relations between these entities (Figure 2). More specifically, CMS Sematrix accepts text documents (primarily scientific or technical) as inputs and then extracts both entities (e.g., health status, change concept, or output) and the significant relationships between and among them using a pipeline of NLP modules. It uses the resulting semantic Web Ontology Language (OWL)/RDF[7] knowledge base to support semantic query and graph visualization. CMS Sematrix is scalable and can process approximately 25,000 documents per day per processing core. It has already processed 8.5 million PubMed abstracts and 1.9 million PubMed Central full articles from 2005 to the present.

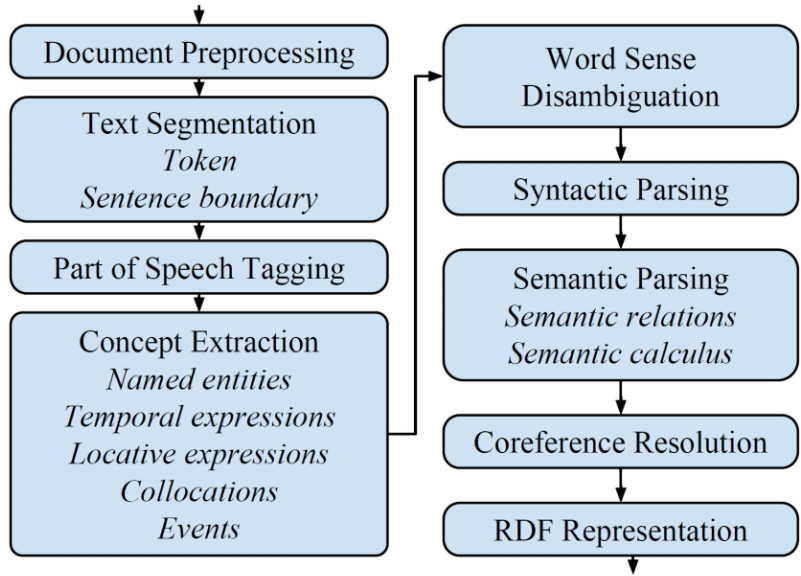

**Figure 1. K-Extractor's Deep NLP Pipeline.** The flow of K-Extractor's NLP pipeline starting from the raw, unstructured document text and ending with the structured RDF triple representation of the extracted concepts and relations.

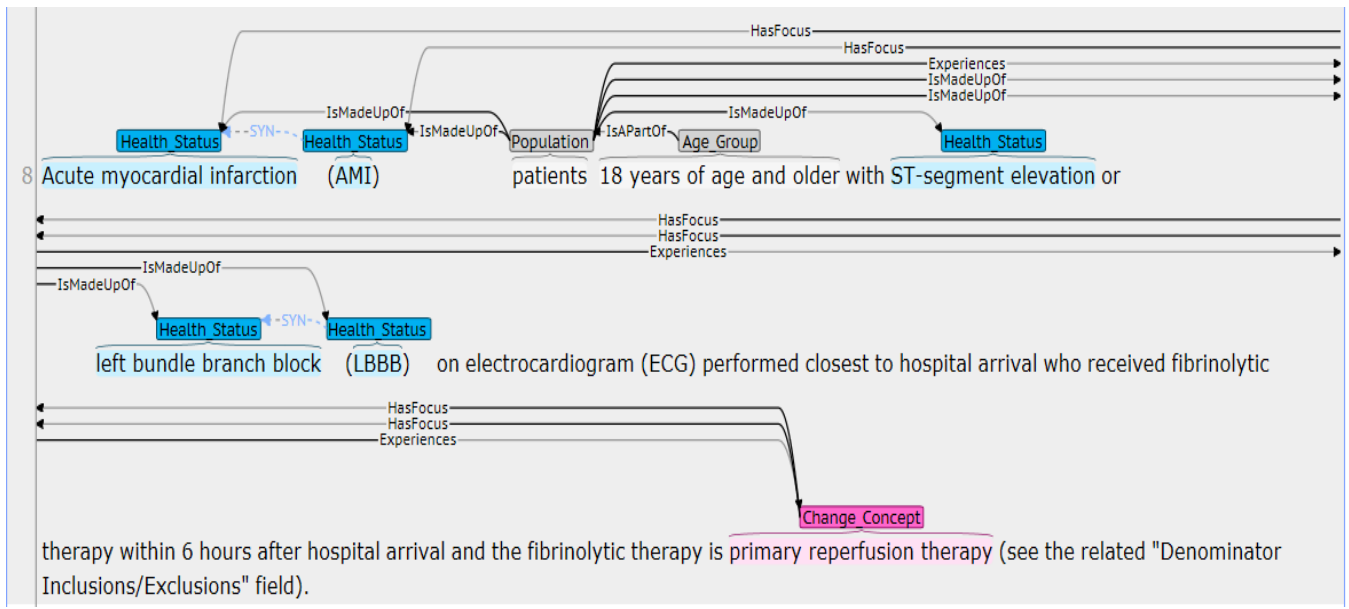

**Figure 2. An example of the entities and semantic relations identified by CMS Sematrix along with their associated ontology classes.** Screenshot of the text of a measure description that has been annotated with the CQM concepts and relations using the brat annotation tool. This section of the text only contains health statuses (AMI, ST-segment elevation, and LBBB), the affected population (patients 18 years and older), and the change concept (primary reperfusion therapy). They are each connected by the appropriate sematic relation.

## Concept Identification

K-Extractor's concept detection methods range from the detection of simple nominal and verbal concepts to more complex named entity and phrasal concepts. The use of a hybrid approach to named entity recognition using machine learning classifiers, cascades of finite-state automatons, and lexicons makes it possible to label more than 80 types of domain independent named entities, including person, organization, and various types of locations, quantities, numerical values, etc.

The finite-state automatons framework uses a pattern-based machine-learning approach and hand-coded rules which allows for a highly customizable and adaptable process for detecting domain relevant concepts including signs, symptoms, disorder, disease, complication, functional status, advanced illness, population, outcome, etc. To learn the lexicon and rules-based models for extracting CQM ontology specific concepts, a set of 65 quality measures and 98 biomedical articles (PubMed Abstracts and PubMed Central Full Articles) were manually annotated using the Brat rapid annotation tool [9], [10] (Figure 2 and see appendix for the list of measures and articles). A random 80:20 split of the manually annotated data was created for training and testing respectively.

## Semantic Relation Identification

In the K-Extractor, semantic relations are instruments used to abstract underlying linguistic relations between concepts. Semantic relations can occur within a word, between words, between phrases, and between sentences. Because semantic relations provide connectivity between concepts, their extraction from text is essential for the ultimate goal of machine text understanding. We use a fixed set of 26 relationships [11] (Table 3), which strike a good balance between too specific and too general. They include the thematic roles proposed by Fillmore [12] and others [13], and the semantic roles in PropBank [14], while also incorporating relationships outside of the verb-argument settings, which highlight key interactions between entities, events, causes, time and space, and others.

**Table 3. K-Extractor's 26 base semantic relations.** The 26 relations in K-Extractor that are used to construct the higher level CQM relations.

| Relation | Definition | Code | Relation | Definition | Code |
|---|---|---|---|---|---|
| Agent(X,Y) | X is the agent of Y; X is proto-typically a person | AGT | Association(X,Y) | Person X is associated with Person Y; the relation is not necessarily kinship | ASO |
| At-Location(X,Y) | X is at-location Y or where X takes place | AT-L | At-Time(X,Y) | X is at-time Y or when X takes place | AT-T |
| Cause(X,Y) | X causes Y | CAU | Experiencer(X,Y) | X is an experiencer of Y; involves cognition and senses | EXP |
| Influence(X,Y) | X caused something to happen to Y | IFL | Instrument(X,Y) | X is an instrument in Y | INS |
| Intent(X,Y) | X is the intent/goal/reason of Y | INT | IS-A(X,Y) | X is a (kind of) Y | ISA |
| Justification(X,Y) | X is the reason or motivation or justification for Y | JST | Kinship(X,Y) | X is a kin of Y; X is related to Y by blood or by marriage | KIN |
| Make(X,Y) | X makes Y | MAK | Manner(X,Y) | X is the manner in which Y happens | MNR |
| Part-Whole(X,Y) | X is a part of Y | PW | Possession(X,Y) | X is a possession of Y; Y owns/has X | POS |
| Property(X,Y) | X is a property/attribute/value of Y | PRO | Purpose(X,Y) | X is the purpose for Y | PRP |
| Quantification(X,Y) | X is a quantification of Y; Y can be an entity or event | QNT | Recipient(X,Y) | X is the recipient of Y; X is an animated entity. | RCP |
| Source(X,Y) | X is the source, origin or previous location of Y | SRC | Stimulus(X,Y) | X is the stimulus of Y; Perceived through senses | STI |
| Synonymy(X,Y) | X is a synonym/name/equal for/to Y | SYN | Theme(X,Y) | X is the theme of Y | THM |
| Topic(X,Y) | X is the topic/focus of cognitive communication Y | TPC | Value(X,Y) | X is the value of Y | VAL |

K-Extractor uses a hybrid approach to semantic parsing. This hybrid approach includes machine learning classifiers for argument pairs identified using syntactic patterns and filtered using extended definitions for our semantic relationships, which describe the possible domain and range information for a relation and impose these semantic restrictions on candidate arguments [11]. Additional modules with specific relational targets are also used.

The below example depicts the conversion of text into a graph by automatically extracting the base semantic relations listed in Table 3 using the K-Extractor:

*"The cfr gene, originally identified in a bovine Staphylococcus sciuri isolate, was found to code for a RNA methyltransferase which modifies the adenine residue at position 2503 in the 23S rRNA and thereby confers resistance not only to oxazolidinones, but also to phenicols, lincosamides, pleuromutilins, and streptogramin A antibiotics."*

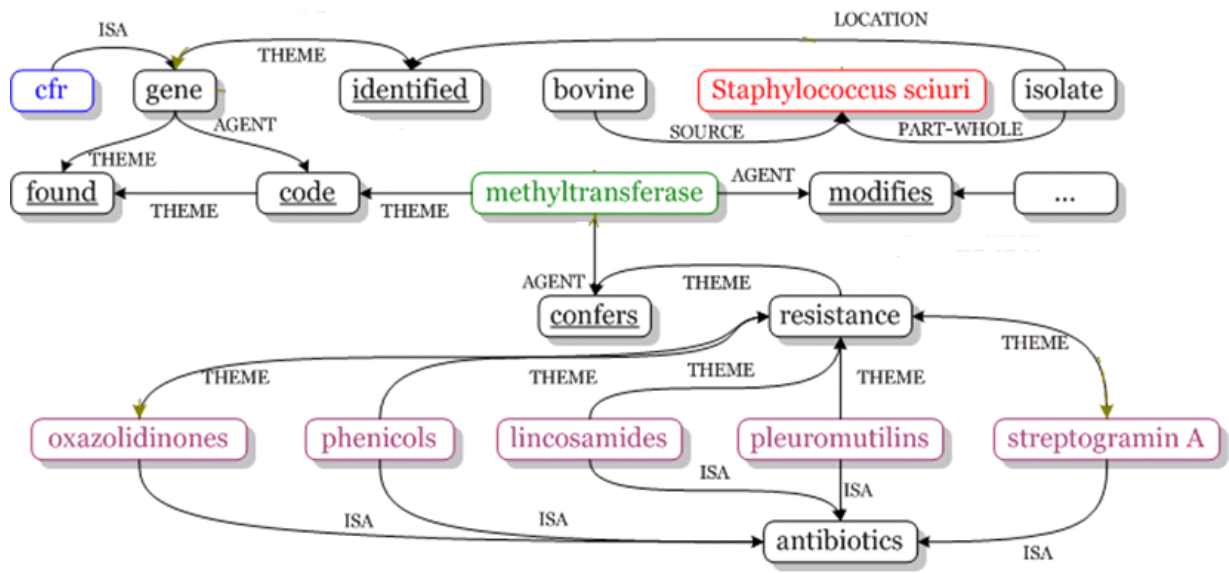

**Figure 3. Example depicting the extraction of base semantic relation from text.** How K-Extractor represents the semantic relations between the entities in the example sentence given above.

The base 26 semantic relations can capture the underlying linguistic relations between concepts in text. However, the base semantic relations do not directly map to the relations defined in domain ontology such as the CQM ontology. K-Extractor provides an easily extensible framework to extract the domain specific relations defined in the domain ontology. K-Extractor uses Semantic Calculus rules (Tatu & Moldovan, 2006) to the extract new types of semantic relations by defining how two or more base semantic relations can be combined. The Semantic Calculus defines axioms for semantic relations $R_0$ that may hold between two concepts $c_1$ and $c_2$, which are linked by two semantic relationships $R_1$ and $R_2$ (not necessarily distinct) that share a third concept $c_3$ as a common argument. More formally:

$$R_1(c_1, c_3) \& R_2(c_3, c_2) \rightarrow R_0(c_1, c_2)$$

For instance, the semantic calculus axiom:

$$\text{Theme}(x, y) \& \text{Cause}(y, z) \rightarrow \text{IsMadeUpOf}(x, z),$$

where *x* is a population concept, z is a health status concept, and y is an experiencing related verb. This axiom can be used to derive new semantic information IsMadeUpOf (patients, hypothyroidism from a sentence such as "The impact of yoga upon female patients suffering from hypothyroidism".)

To automatically learn the semantic calculus axioms required to extract CQM ontology specific semantic relations, the same 65 quality measures and 98 biomedical articles were manually annotated with CQM ontology specific relations. The semantic calculus axioms learning framework used a high recall focus to automatically learn more than 20,000 axioms using the manually annotated examples.

## Knowledge Structures

The CMS Sematrix NLP and knowledge base use a commonly accepted model for knowledge representation known as the Semantic Web. Each piece of knowledge extracted from the text (and tables) in the literature by means of NLP can be stored in a standard and open format known as RDF. The RDF specification represents each statement or assertion as a common data structure, known colloquially as a "triple", that can be thought of as similar to the grammatical notion of "subject [entity] – verb [relationship] – object [entity or value]". Specific (e.g. "named") entities are assigned to classes, or categories, which may be defined in a logical domain ontology (see below). For example, the sentence "John Smith suffers from hypertension" encountered in the text could be represented in pseudo-RDF as something like the example in Table 4.

**Table 4. Example of an RDF triple.** The RDF triple representation of the two entities and single sematic relation from the sentence "John Smith suffers from hypertension."

| RDF | Subject (Domain) | Verb (predicate) | Object (Range) |
|---|---|---|---|
| **Ontology** | Class=PERSON | Property= | Class=DISORDER |
| **Entity** | "John Smith" (specific person) | "suffers from" | "hypertension" |

The RDF triples and associated metadata extracted from each document are stored in a graph database. Knowledge graphs can be constructed from the triples by connecting the like entities or more generally, like concepts per the ontology. This allows for graphs that span specific mentions of an entity within a document or can span documents if desired. For example, in Figure 4 we see the components of a health status dependent quality measure represented as a graph on the measure ontology. The NOT edges represent relationships that are disallowed between the two nodes in the graph.

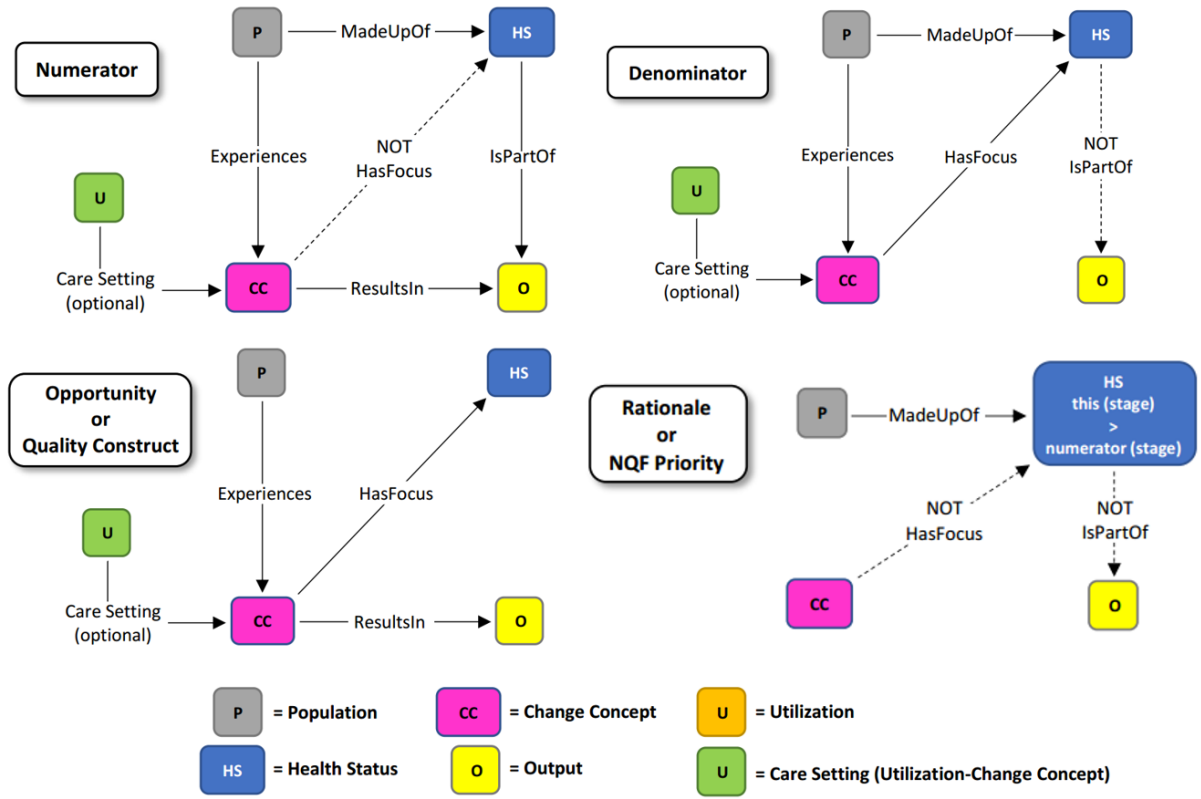

**Figure 4. The 4 components of a health status dependent measure.** The Numerator, Denominator, Opportunity for Improvement, and Rationale components are represented as graphs made up of the five concepts and five sematic relations from the CQM ontology. Care Setting (Utilization-Change Concept) is an optional concept that denotes the setting where the patient experiences the Change Concept.

It is important to note that, in Figure 4, the Health Status that appears in the Numerator can (and often) is different from the Health Status that appears in the three other graphs. However, the Population, Change Concept, and Output are the *same* across all four component graphs.

To see this more concretely, we provide the following example in Table 5 from CMS Measures Inventory Tool (CMIT) measure number 4 titled "Aspirin Prescribed at Discharge".

**Table 5. CQM Concepts Extracted from CMIT 4**. Manually extracted CQM concepts from CMS Measures Inventory Tool measure number 4.

| Measure Description | Population | Denominator Health Status | Change Concept | Numerator Health Status | Output |
|---|---|---|---|---|---|
| **Acute myocardial infarction (AMI) patients who are prescribed aspirin at hospital discharge** | Patients (Age Group: 18 years and older) | Acute myocardial infarction | Aspirin | Mortality | Reduce |

## Matching Publications to Measures

Once the knowledge structures have been extracted from the journal publications, the same procedure is applied to the measure description text. Five knowledge structures (i.e. Keywords, Biomedical Concepts, Biomedical Concept Expansions, Semantic Relations, and CQM Model Semantic Relations) extracted by the K-Extractor from the measure descriptions are then used to create a semantic query. The semantic query consists of 5 different fields/components (one per knowledge structure) and each knowledge structure's field is assigned an importance weight. The process to compute the importance weight is introduced in the Optimizing the Component Weights section. The semantic query is used to match against the same 5 components extracted from the documents index in the publication database with the goal of returning publications that contain relevant information to the measure description. Each publication is then returned with a score denoting its relevancy to the given measure. The overall score utilized by CMS Sematrix is the Lucene Practical Scoring Function [16]:

$$score(q,d) = \frac{\sum_{f \in q} weight(f) \cdot coord(q,d,f) \cdot fieldNorm(f) \left( \sum_{t \in f} value(t) \cdot tf(t,d) \cdot (idf(t))^2 \right)}{\sqrt{\sum_{f \in q} (weight(f))^2 \left( \sum_{t \in f} [value(t) \cdot idf(t)]^2 \right)}},$$

where *q* is a search query, created from processing some inputs (for example, one can be created from a measure XML), *d* is a document in the search index, *f* is a field or component of the score (see below), *t* is a term in the field, *weight(f)* is a weight given to a particular field to boost its importance in the overall score, *coord(q,d,f)* is used to reward documents that contain a higher percentage of the query terms,

*fieldNorm(f)* is the inverse square root of the number of terms in the field, *value(t)* is the value that a term has in the query (typically the number of occurrences of this term in the query input), *tf(t,d)* is the term frequency value for the term *t* in document *d*, and *idf(t)* is the inverse document frequency for term *t* among all documents (the logarithm of the number of documents in the index, divided by the number of documents that contain the term). The full details of the terms making up the Lucene Practical Scoring Function can be found in their documentation [16]. CMS Sematrix uses 5 different components (detailed below) when computing the document score.

As currently implemented, CMS Sematrix searches either the abstract or article component of the content management system for the most relevant documents associated with a single measure. One measure is searched at a time. The system returns the number of documents requested by the user with the top 30 highest overall relevance scores for that measure. The system is not designed to report the five individual scores utilized in the overall score function. It returns the overall, measure-specific relevance score associated with each document searched.

## Individual Score Components

Each of the 5 component scores are defined by various levels of the ontology utilized in CMS Sematrix's natural language processing:

1. **Keywords**: This includes all lemmatized and non-stopword words in an article/measure. Lemmatization is the process of normalizing/generalizing morphological variations of a word to its base form. Example: adults lemmatized to adult, diagnosed lemmatized to diagnose, scanning lemmatized to scan, etc. Stopword removal is the process of removing non-content words such as a, the, and, on, etc.

2. **Biomedical Concepts**: This includes all biomedical concepts present in an article/measure. A concept is a thing mentioned in an article/measure that can span from 1 to n words. Example: high blood pressure, water pressure, video recording, Mediterranean diet, etc. A biomedical concept is a concept that is valid for a particular biomedical domain. Example: for the CMS project, we include all biomedical concepts such as high blood pressure, Mediterranean diet, etc.

3. **Biomedical Concept Expansions**: This includes weighted conceptual expansions of biomedical concepts present in an article/measure. Example: occurrence of high blood pressure in an article/measure will result in concepts such as hypertension, blood pressure, etc. being added (with an appropriated similarity weight) into this field for matching purposes.

4. **Semantic Relations**: This includes semantic relation triples (concept1 semantic-relation-type concept2) that occur between biomedical concepts present in an article/measure. Example: hypertension isCauseOf headache, 120/80 isValueOf blood pressure, etc.

5. **CQM Model Relations**: This includes CQM model specific semantic relation triples that occur between biomedical concepts present in an article/measure. Example: *asian* isAPartOf *patients*, *adult* experiences *mediterranean diet*, etc. The full list of CQM model semantic relations can be found in Table 2.

## Optimizing the Component Weights

To ensure that articles relevant to a given measure are scored higher in the measure search results, the weights for each of the 5 components are optimized in order to maximize a modified mean reciprocal rank (MRR) score [17]. MRR is a statistic for evaluating any process aimed at selecting a best option by ranking the options, ordered by a score. It is calculated as the average of the inverse rank of the best option over multiple executions of the process. The process being evaluated here has multiple correct options. Due to this difference, a modification of the MRR is proposed:

$$v_i = \frac{1}{M_i} \sum_{j=1}^{m_i} \frac{1}{\rho_{ij} - j + 1},$$

where $M_i$ denotes the total number of cited articles associated with the measure of the $i$th search, $m_i$ denotes the number of cited articles in the measure description returned by the $i$th search, and $\rho_{ij}$ denotes the rank (determined by ranking the scores from highest to lowest) of the $j$th cited article returned by the $i$th search.

In order to efficiently determine the optimal set of weights for the 5 components, the Lucene practical scoring function was re-formulated in a way that allows the scores for each component to be run independently from the field/component weights. This means that entire search query does not need to be re-run every time a new set of weights is tested. Briefly, the score for each individual component can be written in terms of weight independent Numerator and Denominator parts so that the total score is computed as follows:

$$score(q,d) = \frac{\sum_{f \in q} weight(f) Numerator(q,d,f)}{\sqrt{\sum_{f \in q} weight(f)^2 Denominator(q,f)}}$$

where $q$ is the current query, $d$ is the given document, *weight(f)* is the weight for the current field (component), and *Numerator(q,d,f)* and *Denominator(q,f)* are combinations of SOLR functions used by the Lucene practical scoring index given above.

To generate the dataset for the weight optimization, queries were run for 65 measures used in the NLP training. Five separate queries were run for each measure in which only 1 component out of the 5 was set to one while the rest were set to zero. For each measure and each component, the Numerator and Denominator parts were computed for each document returned by the query and the top 1,000 documents with the highest numerator parts are returned (since the Denominator part is independent of the document). Thus, for a given set of weights, the Lucene Practical Scoring Function can be computed without re-running the search query to obtain the Numerator and Denominator parts.

One issue that arises when computing the overall score is that, for a given measure, an article can appear in the top 1,000 results for one component and may not appear in the results for the other components. That is, a given document from a particular measure query may not have Numerator and Denominator parts for all 5 fields/components. When computing the combined score across the 5 components for a given document and measure query, if that document is missing a Numerator and Denominator part for a given field, then the Numerator for that document is set to the *minimum* Numerator value found in the 1,000 search results for the field for the given measure. Since the Denominator is independent of the document, it is set to the *same* Denominator value as all other 1,000 search results for the field and the given measure.

To find the optimal component weights, a grid search was performed where the weights for each of the five components varied between 0 and 1 in 0.1 increments results in 115(=161,051) possible weight combinations. The weight combination that maximized the average MRR across the 63 measures was selected. We obtained the best MRR of 0.1098 (for 65 measures) for the following components' weight combination: $W_{Keywords}=0.1$, $W_{Concepts}=0.3$, $W_{Expansion}=0.2$, $W_{Relation}=1.0$, $W_{CQM\_Relation}=0.3$.

### Identifying Relevant Measure Concept Graphs

The Lucene Practical Scoring Function used for scoring a documents' relevancy to a measure is focused around discovering specific terms and relationships within a document. That is, it does not consider the specific structure of the measure concept graphs shown in Figure 4. To assess the degree to which the associated literature provides evidence for the given measure, we developed a procedure for identifying specific, relevant measure concepts in literature associated with each measure by the monthly environmental scan. The goal of this was twofold: (1) to provide a separate verification that the documents returned by the Lucene Practical Scoring Function contain information relevant to a given measure; and (2) to allow measure developers to more quickly and efficiently review environmental scan results.

First, the RDF triples for a given article or abstract are retrieved. As mentioned above, each triple contains the subject text, relationship text, and object text, along with attributes such as: the standardized subject and object text, called subject and object alias text, which collapse instances like hemorrhagic stroke and brain hemorrhage into a single concept by mapping the subject and object text instances to Unified Medical Language System [18] and K-Extractor lexicons (Tatu et al., 2016); the identifier of the document from which the triple was extracted; a measure of the system's confidence in the correctness of the assigned relationship expressed by the triple; and the entity type (class). Below is an example of a triple returned by CMS Sematrix:

| | |
|---|---|
| Document | PMC-4961993 |
| Subject Type | Change Concept |
| Subject Text | Hemodynamic Monitoring |
| Subject Alias Text | Hardiovascular Monitoring |
| Relationship | HasFocus |
| Object Type | Health Status |
| Object Text | Heart Failure |
| Object Alias Text | Heart Failure |

Next, the triples are converted to a graph structure where the nodes are the instances of the concepts extracted from the document (for example, a Health Status node could be Heart Failure) and the edges between the nodes are the semantic relations. All the triples in CQM ontology from given document are combined to form a large "document graph."

## Creating Document Graphs

When constructing the document graphs, it was discovered that there are often instances that appear in the triples that should be merged together. For example, the acronym AMI and the phrase Acute Myocardial Infarction both appear and would be treated as separate nodes in the graph. Not doing this leaves a more disconnected graph as edges that should be associated with just Acute Myocardial Infarction, get separated out among two different nodes. To retrieve the full phrase, these acronyms, any strings with a small number of characters (<=5), are searched in the Text2Knowledeg Acronym Finder database [19]. Additionally, we found that Population and Output tended to vary quite a bit within a given document which also resulted in a disconnected document graph. To remedy this, we converted the text of any tagged Population and Output entities to generic 'Population' and 'Output'.

## Finding Measure Concept Graphs

Next, subgraph matching algorithms are used to identify subgraph-patterns consistent with the "concept maps" in Figure 4, which represent the four basic elements used in the creation of definition of a measure. Essentially, these algorithms enumerate all potential subgraphs of the large document graph in an efficient manner and check them against the aggregate graph pattern to determine whether they are isomorphic (e.g., they have the same node types and relations between them). To perform the subgraph matching, we use the R programming language implementation of the VF2 subgraph isomorphism algorithm[20]. Only the first three concept maps have been used in the graph-matching analysis so far, as the Rationale requires a more sophisticated algorithm than subgraph matching.

## Determining Relevancy of Measure Concept Graphs Found in Documents

The subgraph matching algorithm discussed in the previous section only returns subgraphs that match the pattern of the associated measure concept graph. It does not reveal anything about the relevancy of the instances of Population, Health Status, Change Concept, Output, or Utilization that appear in the subgraph to those of the measure concept graph constructed from the actual measure description. Thus, one could potentially be faced with hundreds of subgraphs that all match the concept graph patterns, but a number of those subgraphs can contain instances of Population, Health Status, Change Concept, Output, or Utilization that are not actually relevant to the current measure. Thus, we then developed a procedure to filter out the non-relevant subgraphs so as to verify that CMS Sematrix is returning documents that contain information that is relevant to the measure being searched.

First off, the concept graphs associated with each measure needed to be extracted from the measure descriptions in order to have a "gold standard" to which potential subgraphs are compared. Initially, this gold standard data set was created by manually reviewing the descriptions for the 65 CQMs to extract the necessary triples to construct the measure concept graphs, but one measure description did not contain enough information for constructing the measure concept graphs. In particular, the instances of Population, Health Status, Change Concept, Output, and Utilization were determined from the measure descriptions (e.g., Table 5).

Each potential measure concept graph that is found by the subgraph matching algorithm in a document is then compared to the manually derived measure concept graphs from measure description in order to determine if there is a match. The methodology for determining a match is as follows: for a given Numerator, Denominator, or Opportunity graph (Figure 4),

1. The text for Health Status, Change Concept, or Utilization nodes in documents and manually derived measure concept graphs are extracted.
2. If there is an exact string match for text in a given node type between the document and manually derived graphs, then that those nodes are said to match.
3. If that does not return a match, then the exact string match is relaxed
    a. Instead, the two strings are compared using a vector model representation of words known as word2vec [21].
    b. The word2vec model utilized was trained using PubMed abstracts from 2005 and used an embedding length of 400.
    c. The distance (1 minus Cosine Similarity) was then used to compare the vector representations of the text from the measure description and document.
    d. If the distance is less than or equal to 0.6, then the nodes are said to match.
4. If there is no exact string match and the distance score is greater than 0.6, then the nodes are said to not match.

A given document graph is considered a match only if every node in the graph is found to be a match to the corresponding nodes of the manually derived concept graph from the measure description.

Lastly, we say that a document returned by CMS Sematrix for a given measure is relevant if and only if it contains at least one (Numerator, Denominator, or Opportunity) graph that matches the corresponding manually derived measure concept graph. However, in results, we also look at the more stringent case where a document is considered relevant if and only if it contains matching Numerator, Denominator, and Opportunity graphs.

# Results

## Validating NLP System

In all the experiments listed in this section, 80% of the annotated data was randomly selected for training the CQM concept and semantic relation extraction modules. The remaining 20% of the annotated examples were used for the testing the quality of the trained modules and computing evaluation results.

Concept lexicons and rules were learnt using the manually annotated examples. The training was focused on maximizing the recall (the fraction of the correct concepts that are successfully identified). Table 6 provides a summary of the number CQM concepts instance examples manually annotated in the test set of quality measures and biomedical articles, and the recall results obtained by the trained NLP models. The models were trained and tested on either only the annotated quality measure data or only the annotated biomedical article data.

**Table 6. CQM Concept Extraction Results.** K-Extractor's performance on extracting the CQM concepts from either the measure text (Quality Measures columns) or the article text (Biomedical Articles columns). In each case, K-Extractor was trained and tested either only on the measure text, or only on the article text.

| Concepts | Quality Measures | | | Biomedical Articles | | |
|---|---|---|---|---|---|---|
| | # Annotated Examples | | Trained NLP Model Recall | # Annotated Examples | | Trained NLP Model Recall |
| | Train Set | Test Set | (on Test Set) | Train Set | Test Set | (on Test Set) |
| **Change Concept** | 1567 | 392 | 86% | 6990 | 1748 | 85% |
| **Health Status** | 2497 | 624 | 92% | 8086 | 2022 | 90% |
| **Output** | 302 | 76 | 76% | 2018 | 504 | 75% |
| **Population** | 1100 | 275 | 95% | 3783 | 946 | 93% |
| **Utilization** | 1056 | 264 | 86% | 2470 | 617 | 85% |

For the final NLP model used in the CMS Sematrix systems, all of the annotated data was used for training.

Table 7 provides a summary of the number CQM semantic relation instance examples manually annotated in the quality measures and biomedical articles, and the recall and precision (the fraction of identified concepts that are correct) results obtained by the trained NLP models. The models were trained and tested on either only the annotated quality measure data, or the annotated measure and biomedical article data. As with the concepts, the final semantic relations extraction model used in the CMS Sematrix systems used all of the annotated data for training.

**Table 7. CQM Semantic Relation Extraction Results.** K-Extractor's performance on extracting the CQM relations from either the measure text, or the measure and article text. In each case, K-Extractor was trained and tested either only on the measure text, or on the measure and article text.

| Source | # Examples Annotated | | Test Results | | | |
| --- | --- | --- | --- | --- | --- | --- |
| | | | Model Trained With Measure Annotations Only | | Model Trained With Measure + Article Annotations | |
| | Train Set | Test Set | Precision | Recall | Precision | Recall |
| **Measure Only** | 6073 | 1518 | 92% | 65% | | |
| **Measure + Articles** | 82536 | 20634 | | | 84% | 62% |

The CMS Sematrix content management system includes abstracts from PubMed from 2007-2018 and full text articles from PubMed Central over the same time period. Additionally, licenses for all articles cited during the development of the core and high impact measures that are NQF endorsed were obtained, and these articles are included in the content management system. Currently, the content management system includes approximately 8.5 million abstracts and over 1.9 million full text articles.

## Validating Results Returned by Sematrix

As mentioned in Methods, a document returned by CMS Sematrix for a given measure is considered *relevant* if and only if it contains at least one measure concept graph (Numerator,

Denominator, or Opportunity) that matches the corresponding manually derived measure concept graph; and is *stringent relevant* if and only if it contains matching Numerator, Denominator, *and* Opportunity graphs. To validate that the documents returned by CMS Sematrix contain information relevant to the given quality measure, we examined the associated top 30 articles for 9 randomly selected measures from the set of 65 CQMs. The 9 randomly selected measures are CMIT 4, 254, 573, 888, 1014, 1241, 1765, 1898, and 2552 (see the corresponding measure descriptions in Appendix). Each article was manually reviewed to determine if it contained information relevant to the associated quality measure.

The results of the manual review were then compared to the results obtained using our automated method to determine *relevant* and *stringent relevant* documents. For the comparison, the results from the manual review were considered to be the "true" relevant documents. Figure 5 shows boxplots of the precision (the fraction of automatically identified relevant documents that were also identified as relevant by the manual procedure) and recall (the fraction of manually identified relevant documents that were also identified as relevant by the automated procedure) scores for *relevant* and *stringent relevant* results aggregated across the 9 measures. Overall, the average precision and recall are 84% and 88%, respectively, both of which indicate that our automated approach can successfully determine relevant documents. In addition, the *stringent relevant* approach would slightly increase the average precision (85%) but causes a large drop in the average recall (56%) which indicates that the *relevant* approach better aligns with the results of the manual review.

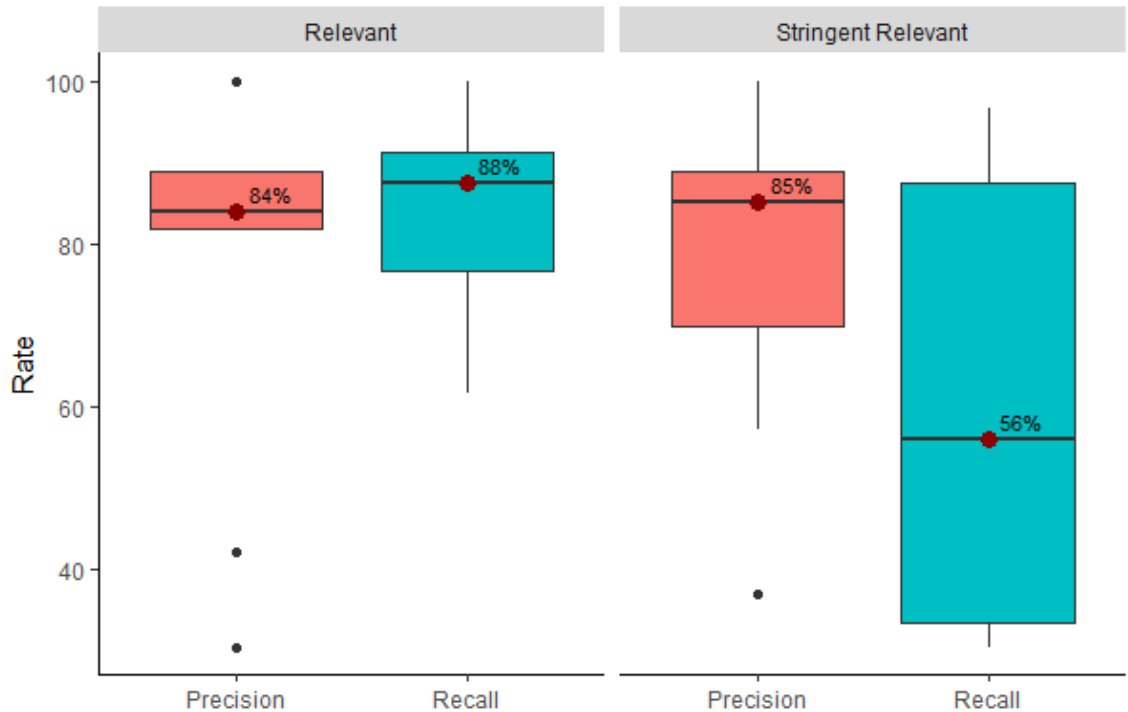

**Figure 5. Comparison of the automated method for determining relevant documents against the manually determined relevant documents for the set of 9 random measures.** Boxplots showing the precision and recall scores for the automated relevancy method using either the *relevant* (left) or *stringent relevant* (right) criteria. The thick horizontal line denotes the median, the lower and upper hinges correspond to the first and third quartiles, the whiskers extend from the hinge to the most extreme value no further than 1.5 × interquartile range from the hinge, and the dots beyond the whiskers are outliers.

To provide an example of the information that our automated relevancy procedure extracts from documents, we reviewed the *relevant* documents returned for measure CMIT 4 (described in Table 5). Table 8 shows an example of the relevant Numerator, Denominator, and Opportunity graphs extracted from the article PMC-4631331 titled "Acute Myocardial Infarction Risk in Patients with Coronary Artery Disease Doubled after Upper Gastrointestinal Tract Bleeding: A Nationwide Nested Case-Control Study" which was deemed *relevant* by our automated procedure. The Denominator and Numerator Health Statuses in the article (acute myocardial infarction) are exactly the same as those extracted from the measure description (see Table 5), while the articles' Change Concept (antiplatelet therapy) is closely related to the Change Concept extracted from the measure description (aspirin). There are several Outputs found in the article graphs as the automated search procedure treats Outputs as a single generic value (see Methods). Nonetheless, the Output extracted from the measure description (reduce) is found among of the list of outputs in the article graphs. Thus, this compact representation very clearly shows that this article contains information relevant to quality measure CMIT 4 and can be utilized by measure developers to quickly summarize and parse the scientific literature.

**Table 8. The relevant Numerator, Denominator, and Opportunity graphs extracted from the article titled "Acute Myocardial Infarction Risk in Patients with Coronary Artery Disease Doubled after Upper Gastrointestinal Tract Bleeding: A Nationwide Nested Case-Control Study" for measure CMIT 4.**

| Population | Denominator Health Status | Change Concept | Numerator Health Status | Output |
|---|---|---|---|---|
| **Patients Cohort** | Acute myocardial infarction | Antiplatelet Therapy | Mortality | Reduce<br>Doubles<br>Internal standard<br>Reduce<br>Was<br>Risk assessment and exposure<br>Prevent |

Next, we used the results from CMIT 4 to investigate the discrepancy between the documents that were determined relevant via manual review and those deemed *relevant* by our automated procedure. Table 9 provides two examples of returned documents for measure CMIT 4 that had different relevancy results from the two different methods.

**Table 9. Examples of the returned documents for measure CMIT 4 that had different relevancy results for the automated method and the manual examination.**

| PubMed Central ID | Article Title | Relevancy | |
|---|---|---|---|
| | | Manually Examined | Automated Method |
| PMC-539261 | Differences in access to coronary care unit among patients with acute myocardial infarction in Rome: old, ill, and poor people hold the burden of | No | Yes |
| PMC-5862020 | Chronic obstructive pulmonary disease and acute myocardial infarction: effects on presentation, management, and outcomes | Yes | No |

For PMC-539261, although the term "aspirin" is used once in the discussion section, in context the term does not significantly contribute to evidence supporting the objective of the article. The term "aspirin" may not be a suitable Change Concept when considering the article as a whole; rather, "standard optimal coronary care" is the appropriate Change Concept, as stated in the abstract. On the other hand, although PMC-5862020 is not strictly focused on acute myocardial infarction (AMI) as a Health Status, the article demonstrates important correlations in approaches to therapy between AMI and Chronic Obstructive Pulmonary Disease (COPD) that may be useful to providers, such as the use of antiplatelet therapy (i.e., the focus of CMIT-4). To emphasize this point, the article presents in evidence in a tabular format that COPD is an important factor in whether AMI patients receive aspirin at discharge. However, the table uses complex formatting (i.e., blank cells; multiple lines of data per cell) that may not be easily processed by Sematrix. Improvement in the way Sematrix digests tables, particularly tables with complex formatting, should improve relevancy scores for articles that use tables to present findings.

Lastly, we looked at the number of articles that are deemed *relevant* and *stringent relevant* by our automated procedure for the top-30 documents returned by CMS Sematrix for each of the 65 measures except for CMIT 967 (the measure description of CMIT 967 did not include enough information for manually annotated the gold standard graphs). The results are shown in Figure 6. We found on average roughly 72% of the articles returned by CMS Sematrix for a given measure contain information relevant to the measure description (i.e., ~21 out of the 30 returned documents). However, there were a few measures where CMS Sematrix did not return any *relevant* documents. For example, CMIT 284 did not appear to have any relevant documents according to our automated procedure. This is most likely due to the fact that the measure description was updated after we extracted the "gold standard" measure concept graphs (see Methods). We also found that CMS Sematrix had relatively poor performance for CMIT 78, 80, 86, and 89 in terms of number of relevant documents returned, which is likely caused by the fact that their measure descriptions did not provide enough information to extract the precise measure concepts

(e.g., the Change Concept was not specified). Thus, a very general change concept (e.g., quality improvement) was inferred making it difficult to match relevant documents.

**Figure 6. Relevancy results across the entire set of 65 CQMs using the automated method.** The bar plot shows the number of documents (out of the 30 returned search results) determined *relevant* (red) or *stringent relevant* (red) by our automated method.

# Conclusion

To effectively evaluate the quality of health care, the developers of clinical quality measures face the arduous task of scanning the biomedical literature each month in order to ensure that the evidence supporting each of their roughly 2,000 CQMs is timely and complete. In this work, we have detailed our tool CMS Sematrix which is aimed at reducing the burden placed on measure developers by effectively automating the knowledge discovery process of the monthly scans. CMS Sematrix contains three major components: (1) A quality measure ontology to describe high-level knowledge constructs contained in CQM; (2) a NLP system to extract concepts and relations that correspond to the ontology from text; and (3) a graphical database to store the concepts and relations extracted from text as RDF triples that can be queried to deduce measure components within documents. We have shown that the NLP component of CMS Sematrix was able to correctly identify CQM concepts with an average recall score of 87% for measure descriptions and 86% for articles. In addition, CMS Sematrix achieved overall precision and recall scores of 84% and 62% when extracting concept relations. We then conducted an environmental scan of the PubMed and PubMed Central abstracts and articles using a set of 65 CQMs. For the 9 measures selected for manual review, our automated procedure for determining relevant documents obtained average precision and recall scores of 84% and 88%. Running this procedure on the full set of

65 CQMs, we found that on average roughly 72% of the articles returned by CMS Sematrix for a given measure contain information relevant to the measure description using our June 2018 environmental scan data.

CMS Sematrix is able to identify articles published in the clinical and health services literature that contain information relevant to a given CQM. In practice, CMS Sematrix can reduce the time-consuming burden of the CMS monthly environmental scans and allow measure developers to quickly and accurately design CQM to track outcomes in order to improve the national healthcare system.

## Acknowledgements

Support for this work was provided in part by the Centers for Medicare and Medicaid Services under task order HHSM-500-T0001 and contract HHSM-500-2013-13005I and Battelle Memorial Institute.

# Appendix

Table A.1 shows the CMIT ID and measure description for the list of 65 quality measures.

*Table A.1. The set of 65 quality measures.*

| CMIT ID | Measure Description |
|---|---|
| 4 | Aspirin prescribed at discharge for AMI |
| 13 | Acute myocardial infarction (AMI): percent of patients with ST-segment elevation or LBBB on the ECG closest to arrival time receiving fibrinolytic therapy during the hospital stay and having a time from hospital arrival to fibrinolysis of 30 minutes or less. |
| 16 | Acute myocardial infarction (AMI): percent of patients with ST-segment elevation or LBBB on the ECG closest to arrival time receiving primary PCI during the hospital stay with a time from hospital arrival to PCI of 90 minutes or less. |
| 78 | Heart failure (HF): hospital 30-day, all-cause, unplanned risk-standardized readmission rate (RSRR) following HF hospitalization. |
| 80 | Acute myocardial infarction (AMI): hospital 30-day, all-cause, unplanned risk-standardized readmission rate (RSRR) following AMI hospitalization. |
| 86 | Acute myocardial infarction (AMI): hospital 30-day, all-cause, risk-standardized mortality rate (RSMR) following AMI hospitalization. |
| 89 | Heart failure (HF): hospital 30-day, all-cause, risk-standardized mortality rate (RSMR) following HF hospitalization. |
| 128 | Acute myocardial infarction (AMI): percentage of ED patients with AMI and ST-segment elevation on the ECG closest to arrival time receiving fibrinolytic therapy during the ED stay and having a time from ED arrival to fibrinolysis of 30 minutes or less. |
| 130 | Acute myocardial infarction (AMI): median time to transfer to another facility for acute coronary intervention. |
| 233 | Chronic stable coronary artery disease: percentage of patients aged 18 years and older with a diagnosis of coronary artery disease seen within a 12 month period who also have prior MI or a current or prior LVEF less than 40% who were prescribed beta-blocker therapy. |
| 243 | Eye care: percentage of patients aged 18 years and older with a diagnosis of primary open-angle glaucoma (POAG) who have an optic nerve head evaluation during one or more office visits within 12 months. |
| 254 | Eye care: percentage of patients aged 18 years and older with a diagnosis of diabetic retinopathy who had a dilated macular or fundus exam performed with documented communication to the physician who manages the on-going care of the patient with diabetes mellitus regarding the findings of the macular or fundus exam at least once within 12 months. |
| 267 | Osteoporosis: percentage of patients aged 50 years and older treated for a hip, spine or distal radial fracture with documentation of communication with the physician managing the patient's on-going care that a fracture occurred and that the patient was or should be tested or treated for osteoporosis. |
| 280 | Stroke and Stroke Rehabilitation: Discharged on Antithrombotic Therapy |
| 284 | Stroke and Stroke Rehabilitation: Anticoagulant Therapy Prescribed for Atrial Fibrillation (AF) at Discharge |
| 291 | Osteoporosis: percentage of women 65 to 85 years of age who have documentation in their medical record of having received a central dual-energy X-ray absorptiometry (DXA) test of their hip or spine. |
| 326 | Chronic obstructive pulmonary disease (COPD): percentage of patients aged 18 years and older who had a spirometry evaluation results documented at least annually. |
| 328 | Percentage of patients aged 18 years and older with a diagnosis of COPD (FEV1/FVC < 70%) and who have an FEV1 less than 60% predicted and have symptoms who were prescribed an long-acting inhaled bronchodilator. |
| 369 | Oncology: percentage of female patients aged 18 years and older with Stage IC through IIIC, estrogen receptor (ER) or progesterone receptor (PR) positive breast cancer who were prescribed tamoxifen or aromatase inhibitor (AI) during the 12 month reporting period. |
| 372 | Oncology: percentage of patients aged 18 through 80 years with American Joint Committee on Cancer (AJCC) Stage III colon cancer who are referred for adjuvant chemotherapy, prescribed adjuvant chemotherapy, or have previously received adjuvant chemotherapy within the 12 month reporting period. |
| 420 | Pathology: percentage of colon and rectum cancer resection pathology reports that include the pT category (primary tumor), the pN category (regional lymph nodes) and the histologic grade. |
| 433 | Disease-modifying anti-rheumatic drug therapy for rheumatoid arthritis: percentage of members who were diagnosed with rheumatoid arthritis and who were dispensed at least one ambulatory prescription for a disease modifying anti-rheumatic drug (DMARD). |
| 451 | Colorectal cancer screening: percentage of members 50 to 75 years of age who had appropriate screening for colorectal cancer. |
| 461 | Comprehensive diabetes care: percentage of members 18 to 75 years of age with diabetes (type 1 and type 2) who had an eye exam (retinal) performed. |
| 464 | Chronic stable coronary artery disease: percentage of patients aged 18 years and older with a diagnosis of coronary artery disease seen within a 12 month period who also have diabetes or a current of prior LVEF less than 40% who were prescribed ACE inhibitor or ARB therapy. |
| 479 | End stage renal disease (ESRD): percentage of patient-months of pediatric (less than 18 years) in-center hemodialysis patients (irrespective of frequency of dialysis) with documented monthly nPCR measurements. |
| 496 | Diabetes mellitus: percentage of patients aged 18 years and older with a diagnosis of diabetes mellitus who had a lower extremity neurological exam performed at least once within 12 months. |
| 499 | Diabetes mellitus: percentage of patients aged 18 years and older with a diagnosis of diabetes mellitus who were evaluated for proper footwear and sizing at least once within 12 months. |
| 533 | Eye care: percentage of patients aged 18 years and older with a diagnosis of primary open-angle glaucoma whose glaucoma treatment has not failed (the most recent IOP was reduced by at least 15% from the pre-intervention level) OR if the most recent IOP was not reduced by at least 15% from the pre-intervention level a plan of care was documented within 12 months. |
| 573 | Oncology: percentage of patients, regardless of age, with a diagnosis of breast, rectal, pancreatic or lung cancer receiving 3D conformal radiation therapy who had documentation in medical record that radiation dose limits to normal tissues were established prior to the initiation of a course of 3D conformal radiation for a minimum of two tissues. |
| 650 | Endoscopy and polyp surveillance: percentage of patients aged 18 years and older receiving a surveillance colonoscopy, with a history of a prior colonic polyp in previous colonoscopy findings who had a follow-up interval of 3 or more years since their last colonoscopy documented in the colonoscopy report. |
| 662 | Cataracts: 20/40 or Better Visual Acuity within 90 Days Following Cataract Surgery |
| 665 | Cataracts: Complications within 30 Days Following Cataract Surgery Requiring Additional Surgical Procedures |
| 844 | Total hip arthroplasty (THA) and/or total knee arthroplasty (TKA): hospital-level risk-standardized complication rate (RSCR) following elective primary THA and/or TKA. |
| 859 | Stroke: percent of ischemic stroke patients prescribed antithrombotic therapy at hospital discharge. |
| 861 | Stroke: percent of ischemic stroke patients with atrial fibrillation/flutter who are prescribed anticoagulation therapy at hospital discharge. |
| 865 | Stroke: percent of ischemic stroke patients administered antithrombotic therapy by the end of hospital day 2 |
| 867 | Stroke: percent of ischemic stroke patients who are prescribed a statin medication at hospital discharge |
| 888 | Stroke: percent of ischemic or hemorrhagic stroke patients who were assessed for rehabilitation services. |
| 889 | Venous thromboembolism (VTE): percent of patients who received VTE prophylaxis or have documentation why no VTE prophylaxis was given the day of or the day after initial admission (or transfer) to the ICU or surgery end date for surgeries that start the day of or the day after ICU admission (or transfer). |
| 918 | Emergency department (ED): percentage of ED acute ischemic stroke or hemorrhagic stroke patients who arrive at the ED within 2 hours of the onset of symptoms who have a head CT or MRI scan performed during the stay and having a time from ED arrival to interpretation of the head CT or MRI scan within 45 minutes of arrival. |
| 967 | In-center hemodialysis patients' satisfaction with care: in-center hemodialysis patients' overall ratings of their dialysis center. |
| 1014 | End stage renal disease (ESRD): percentage of adult dialysis patients with a 3-month rolling average of total uncorrected calcium (serum or plasma) greater than 10.2 mg/dL (hypercalcemia). |
| 1049 | Cataracts: percentage of patients aged 18 years and older in sample who had cataract surgery and had improvement in visual function achieved within 90 days following the cataract surgery, based on completing a pre-operative and postoperative visual function survey. |
| 1070 | Cardiac rehabilitation: percentage of patients in an outpatient clinical practice who have had a qualifying event/diagnosis during the previous 12 months, who have been referred to an outpatient cardiac rehabilitation program. |
| 1117 | Rate of Endovascular Aneurysm Repair (EVAR) of Small or Moderate Non-Ruptured Infrarenal Abdominal Aortic Aneurysms (AAA) Who Are Discharged Alive |
| 1241 | Diabetes mellitus care: percentage of patients 18 to 75 years of age who had a diagnosis of type 1 or type 2 diabetes and whose diabetes was optimally managed during the measurement period. |
| 1246 | Controlling high blood pressure: percentage of members 18 to 85 years of age who had a diagnosis of hypertension (HTN) and whose BP was adequately controlled during the measurement year, based on age/condition-specific criteria. |
| 1275 | Preventive care and screening: percentage of patients aged 18 years and older who were screened for tobacco use one or more times within 24 months AND who received cessation counseling intervention if identified as a tobacco user. |
| 1334 | End stage renal disease (ESRD): percentage of all patient months for adult patients (≥ 18) whose delivered peritoneal dialysis dose was a weekly Kt/Vurea ≥ 1.7 (dialytic + residual). |
| 1336 | End stage renal disease (ESRD): percentage of patient months for patients on maintenance hemodialysis during the last HD treatment of month using an autogenous AV fistula. |
| 1338 | End stage renal disease (ESRD): percentage of patient months on maintenance hemodialysis during the last HD treatment of month with a chronic catheter continuously for 90 days or longer prior to the last hemodialysis session. |
| 1349 | End stage renal disease (ESRD): percentage of patient months for all pediatric ( < 18 years old) in-center hemodialysis patients in which the delivered dose of hemodialysis (calculated from the last measurement of the month using the UKM or Daugirdas II formula) was spKt/V > 1.2. |
| 1367 | Imaging efficiency: percentage of stress echocardiography, SPECT MPI, or stress MRI studies performed at a hospital outpatient facility in the 30 days prior to an ambulatory low-risk, non-cardiac surgery performed anywhere. |
| 1404 | Comprehensive diabetes care: percentage of members 18 to 75 years of age with diabetes (type 1 and type 2) whose most recent hemoglobin A1c (HbA1c) level is greater than 9.0% (poorly controlled). |
| 1406 | Diabetes: Medical Attention for Nephropathy |
| 1437 | Venous Thromboembolism Prophylaxis |
| 1455 | Chronic obstructive pulmonary disease (COPD): hospital 30-day, all-cause, risk-standardized readmission rate following acute exacerbation of COPD hospitalization. |
| 1765 | Percentage of patients aged 18 years and older with a diagnosis of nonvalvular atrial fibrillation (AF) or atrial flutter whose assessment of the specified thromboembolic risk factors indicate one or more high-risk factors or more than one moderate risk factor, as determined by CHADS2 risk stratification, who are prescribed warfarin OR another oral anticoagulant drug that is FDA approved for the prevention of thromboembolism |
| 1898 | Pediatric kidney disease: percentage of calendar months within a 12-month period during which patients aged 17 years and younger with a diagnosis of ESRD receiving hemodialysis or peritoneal dialysis have a hemoglobin level less than 10 g/dL. |
| 1926 | Prostate cancer: percentage of patients, regardless of age, with a diagnosis of prostate cancer at high or very high risk of recurrence, receiving external beam radiotherapy to the prostate who were prescribed adjuvant hormonal therapy (GnRH agonist or antagonist). |

| 1930 | Chronic obstructive pulmonary disease (COPD): hospital 30-day, all-cause, risk-standardized mortality rate following acute exacerbation of COPD. |
|---|---|
| 2331 | Pathology: percentage of breast cancer resection pathology reports that include the pT category (primary tumor), the pN category (regional lymph nodes) and the histologic grade. |
| 2343 | Prostate cancer: percentage of patients, regardless of age, with a diagnosis of prostate cancer at low risk of recurrence receiving interstitial prostate brachytherapy, OR external beam radiotherapy to the prostate, OR radical prostatectomy, OR cryotherapy who did not have a bone scan performed at any time since diagnosis of prostate cancer. |
| 2552 | Osteoporosis management in women who had a fracture: percentage of women 67 to 85 years of age who suffered a fracture and who had either a bone mineral density (BMD) test or prescription for a drug to treat osteoporosis in the six months after the fracture. |

**Table A.2. The list of returned documents for measure CMIT 4 with manually examined and MIF analysis results.**

| Article Title | Validation Score | Relevancy | |
|---|---|---|---|
| | | Manually Examined | MIF Analysis |
| Acute Myocardial Infarction Risk in Patients with Coronary Artery Disease Doubled after Upper Gastrointestinal Tract Bleeding: A Nationwide Nested Case-Control Study | 13 | Yes | Yes |
| Single and dual antiplatelet therapy in elderly patients of medically managed myocardial infarction | 13 | Yes | Yes |
| Reducing mortality in sepsis: new directions | 13 | Yes | Yes |
| The impact of cardiac and noncardiac comorbidities on the short-term outcomes of patients hospitalized with acute myocardial infarction: a population-based perspective | 13 | Yes | Yes |
| Differences in access to coronary care unit among patients with acute myocardial infarction in Rome: old, ill, and poor people hold the burden of inefficiency | 13 | No | Yes |
| Gender Differences in Presentation, Management, and In-Hospital Outcomes for Patients with AMI in a Lower-Middle Income Country: Evidence from Egypt | 13 | Yes | Yes |
| Long-term prognosis of diabetic patients with acute myocardial infarction in the era of acute revascularization | 13 | Yes | Yes |
| Educational level and 30-day outcomes after hospitalization for acute myocardial infarction in Italy | 13 | No | Yes |

| Article Title | Validation Score | Relevancy | |
|---|---|---|---|
| | | Manually Examined | MIF Analysis |
| Regional Differences in Treatment Frequency and Case-Fatality Rates in Korean Patients With Acute Myocardial Infarction Using the Korea National Health Insurance Claims Database: Findings of a Large Retrospective Cohort Study | 13 | Yes | Yes |
| Geographic variation in the treatment of acute myocardial infarction: the Cooperative Cardiovascular Project | 13 | Yes | Yes |
| ACC/AHA 2008 performance measures for adults with ST-elevation and non-ST-elevation myocardial infarction: a Report of the American College of Cardiology/American Heart | 10 | Yes | Yes |
| Differences between acute myocardial infarction and unstable angina: a longitudinal cohort study reporting findings from the Register of Information and Knowledge about Swedish | 10 | Yes | Yes |
| Antithrombotic therapy practices in US hospitals in an era of practice guidelines | 10 | Yes | Yes |
| Trends in early aspirin use among patients with acute myocardial infarction in China, 2001-2011: the China PEACE-Retrospective AMI study. | 10 | Yes | Yes |
| Rural hospital emergency department quality measures: aggregate data report | 10 | Yes | Yes |
| Trends in the Incidence and Management of Acute Myocardial Infarction From 1999 to 2008: Get With the Guidelines Performance Measures in Taiwan | 10 | Yes | Yes |
| Clinical characteristics and improvement of the guideline-based management of acute myocardial infarction in China: a national retrospective analysis | 10 | Yes | Yes |
| Resource use and quality of care for Medicare patients with acute myocardial infarction in Maryland and the District of Columbia: analysis of data from the Cooperative | 10 | Yes | Yes |
| A short-term risk-benefit analysis of occasional and regular use of low-dose aspirin in primary prevention of vascular diseases: a nationwide population-based study | 10 | Yes | Yes |
| Intra-aortic balloon counterpulsation pump therapy: a critical appraisal of the evidence for patients with acute myocardial infarction | 3 | No | Yes |
| Declining mortality following acute myocardial infarction in the Department of Veterans Affairs Health Care System | 3 | Yes | Yes |

| Article Title | Validation Score | Relevancy | |
|---|---|---|---|
| | | Manually Examined | MIF Analysis |
| French Registry on Acute ST-elevation and non ST-elevation Myocardial Infarction 2010. FAST-MI 2010 | 1 | Yes | Yes |
| ACC/AHA 2008 performance measures for adults with ST-elevation and non-ST-elevation myocardial infarction: a report of the American College of Cardiology/American Heart Association Task Force on Performance Measures (Writing Committee to Develop Performance Measures for ST-Elevation and Non-ST-Elevation Myocardial Infarction) Developed in Collaboration With the American Academy of Family Physicians and American College of Emergency Physicians Endorsed by the American Association of Cardiovascular and Pulmo | 1 | Yes | Yes |
| Initiation of and long-term adherence to secondary preventive drugs after acute myocardial infarction | 1 | Yes | Yes |
| Endothelial Progenitor Cells Predict Cardiovascular Events after Atherothrombotic Stroke and Acute Myocardial Infarction. A PROCELL Substudy | 1 | Yes | Yes |
| Chronic obstructive pulmonary disease and acute myocardial infarction: effects on presentation, management, and outcomes | 0 | Yes | No |
| Combined metallic osteosynthesis in fractures of the pelvis | 0 | No | No |
| Synergistic effects of cardiac resynchronization therapy and drug up-titration in heart failure: is this enough | 0 | No | No |
| The association between emergency department crowding and hospital performance on antibiotic timing for pneumonia and percutaneous intervention for myocardial infarction | 0 | No | No |
| Increase in the proportion of patients hospitalized with acute myocardial infarction with do-not-resuscitate orders already in place between 2001 and 2007: a nonconcurrent prospective study | 0 | Yes | No |